\documentclass[aps,pre,twocolumn,superscriptaddress,floatfix]{revtex4}
\usepackage[english]{babel}
\usepackage{graphicx}
\usepackage[dvips,unicode,colorlinks,linkcolor=blue,citecolor=blue,urlcolor=blue]{hyperref}
\begin{document}
\title{Eigenmodes and resonance vibrations of 2D nanomembranes --  Graphene and hexagonal boron-nitride}

\author{Alexander V. Savin}
\affiliation{ N.N. Semenov Federal Research Center for Chemical Physics,
Russian Academy of Sciences (FRCCP RAS), Moscow, 119991, Russia}

\begin{abstract}
Natural and resonant oscillations of suspended circular graphene and hexagonal boron nitride
(h-BN) membranes (single-layer sheets lying on a flat substrate having a circular hole of radius $R$)
have been simulated using full-atomic models.
Substrates formed by flat surfaces of graphite and h-BN crystal, hexagonal ice, silicon carbide 6H-SiC
and nickel surface (111) have been used.
The presence of the substrate leads to the forming of a gap at the bottom of the frequency spectrum
of transversal vibrations of the sheet.
The frequencies of natural oscillations of the membrane (oscillations localized on the suspended
section of the sheet) always lie in this gap, and the frequencies of oscillations decrease
by increasing radius of the membrane as $(R+R_i)^{-2}$ with nonezero effective increase of radius $R_i>0$.
The modeling of the sheet dynamics has shown that small periodic transversal displacements
of the substrate lead to resonant vibrations of the membranes at frequencies close to 
eigenfrequencies of nodeless vibrations of membranes with a circular symmetry.
The energy distribution of resonant vibrations of the membrane has a circular symmetry
and several nodal circles, whose number $i$ coincides with the number of the resonant frequency.
The frequencies of the resonances decrease by increasing the radius of the membrane as
$(R+R_i)^{\alpha_i}$ with exponent $\alpha_i<2$.
The lower rate of resonance frequency decrease is caused by the anharmonicity
of membrane vibrations.
\end{abstract}


\maketitle

\section{Introduction}

\begin{figure}[bt]
\begin{center}
\includegraphics[angle=0, width=1.0\linewidth]{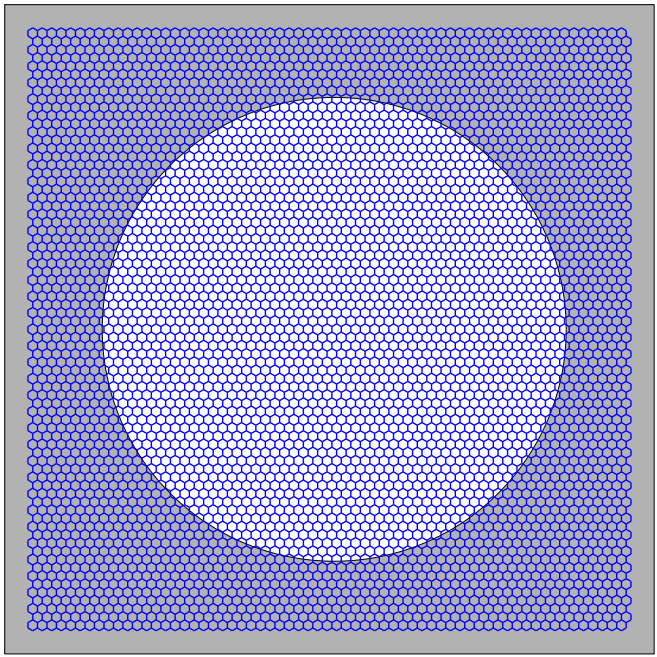}
\end{center}
\caption{\label{fig01}\protect
The rectangular sheet of graphene lying on a flat substrate having a circular hole in its center.
The sheet consists of 9470 carbon atoms and has the shape of a rectangle of size $15.6\times15.6$~nm$^2$.
The radius of the hole in the substrate $R=6$~nm.
}
\end{figure}

Being a nanosized polymorph of carbon, graphene attracts increased attention of researchers
due to its unique physical properties \cite{geim07,soldano10}.
The remarkable properties of graphene have enabled the exploitation of graphene for the development
of nano-electro-mechanical system (NEMS) such as nanoresonators \cite{bunch07,eom11}.
The vibrational properties of graphene play an important role in
analysis and design of graphene-based sensors and resonators.
The aim of this work is to simulate the eigenmodes and resonant vibrations of
suspended circular graphene (G) and hexagonal boron nitride (h-BN) membranes.
Such 2D membranes are formed as single-layer G and h-BN sheets lying on a flat
substrate with a circular hole -- see Fig.~\ref{fig01}.
These one atom thick membranes can be used as highly efficient nanomechanical resonators
\cite{zande10,barton11,barton11a,guttinger17,verbiest18} and as extraordinary sensitive detectors of mass,
force and pressure \cite{bunch08,smith16,akinwande17}.
\begin{table*}[tb]
\caption{
Parameters of the LJ potential (\ref{f1}) for various pairs of interacting atoms.
\label{tab1}}
\begin{tabular}{c|c|cccccccccccc}
\hline
\hline

~                 & \cite{setton96}~CC  &  \cite{rappe92}~CC  &  CH      &  CO      & CSi     & NO      &
NH      & NC      &  NSi       & BO       & BH      &  BC  & BSi \\
$\epsilon_0$ (meV) &  2.76   &  4.56      &  2.95 &  3.44 & 8.92 & 2.78 &
2.38 &  3.69&  7.24   &  4.51 & 3.86 &  5.94 & 11.66 \\
$r_0$ (\AA)       &  ~3.809~        &  ~3.851        &  3.369   &  3.676   & 4.073   & 3.579   &
3.25   & 3.754   &  3.965     &  3.78   & 3.433 & 3.965  & 4.188\\
\hline
\hline
\end{tabular}
\end{table*}

For analysis of vibrations of such membranes continuum models in which
a sheet of graphene is considered as continuous thin plate or thin shell
\cite{atalaya2008,dai12,ghaffari18,shi19}  are usually used.
In this paper we will use discrete (full-atomic) models that take into account the
hexagonal structure of the sheets.
As substrate, we consider the plane surfaces of an ideal graphite and h-BN crystals,
hexagonal ice I$_h$, silicon carbide 6H-SiC and the surface (111) of Nickel crystal.

\section{Model}
To calculate the interaction energy of a sheet (graphene and h-BN) with a flat substrate
the sheet has been placed parallel to the substrate surface.
The interaction potential of each atom belonging to the sheet
with the substrate $W(h)$ can be found  as the function of the distance to the substrate plane $h$,
as a sum of its interaction energies with the substrate atoms. The interaction
of pairs of atoms has been described by the Lennard-Jones (LJ) potential (6,12):
\begin{equation}
V_{LJ}(r)=\epsilon_0[(r_0/r)^{12}-2(r_0/r)^6], \label{f1}
\end{equation}
where $\epsilon_0$ is the binding energy and $r_0$ is the bond length.
To find the interaction energy of graphene with the crystalline graphite surface,
we used the potential parameters taken from \cite{setton96} and, for other
substrates, from \cite{rappe92}. Table \ref{tab1} shows the parameters of LJ
potential (\ref{f1}) for various atomic pairs.

The calculations have been made for the 2.0$\times$ 1.8~nm$^2$ graphene (h-BN) sheet
consisting of 160 carbon (boron and nitride) atoms,
which is arranged in parallel to the crystal surface at distance $h$.
At each value of distance $h$, the energy was averaged over the shifts along substrate surface
and, then, normalized on the number of atoms in the graphene (h-BN) sheet.
As a result, we obtained the dependence of the interaction energy of one atom of the sheet with the
substrate on its distance from substrate plane $W(h)$.
The calculations showed that the interaction energy with the substrate $W(h)$ can be described
with a high accuracy by the Lennard-Jones potential ($k,l$):
\begin{equation}
W(h)=\epsilon_1[k(h_1/h)^l-l(h_1/h)^k]/(l-k), \label{f2}
\end{equation}
where power $l>k$. Potential (\ref{f2}) has the minimum $W(h_1)=-\epsilon_1$
($\epsilon_1$ is the binding energy of the atom with substrate).
The stiffness of interaction with the substrate is $K_1=W''(h_1)=\epsilon_1 lk/h_1^2$.
Table \ref{tab2} presents the parameters of LJ potential (\ref{f2}) for graphene
and h-BN sheet on various substrates.
\begin{table}[tb]
\caption{
Parameters of $(k,l)$ LJ potential (\ref{f2}) for graphene and h-BN sheets on various substrates.
\label{tab2}
}
\begin{tabular}{c|ccccc}
\hline
\hline
~                     & ~$\epsilon_1$~(eV)~&$h_1$~(\AA) &  $l$  & $k$ &   $K_1$ (N/m) \\
Graphene -- Ice I$_h$ &   0.029           &  3.005        &  10   & 3.5 &   1.80  \\
Graphene -- Graphite  &   0.052           &  3.37         &  10   & 3.75&   2.75  \\
Graphene -- 6H-SiC    &   0.073           &  4.19         &  17   & 3.75&   4.24  \\
Graphene -- h-BN      &   0.0903          &  3.46         &  10   & 3.75&   4.53  \\
h-BN -- Ice I$_h$     &   0.0304          &  3.04         &  10   & 3.5 &   1.81  \\
h-BN -- 6H-SiC        &   0.0803          &  4.20         &  17   & ~3.75~& 4.50 \\
\hline
\hline
\end{tabular}
\end{table}

When graphene is located on the surface (111) of crystalline nickel, a stronger
chemical interaction of carbon atoms with the atoms of the substrate occurs
(hybridization of the metal $d$-band with graphene $\pi$-states and charge transfer from
the metal to graphene).
As a result of the interaction of a graphene sheet with a crystal surface a gap of the magnitude
$\omega_0=240$~cm$^{-1}$ appears at the bottom of the frequency spectrum of transversal
oscillations of the sheet~\cite{dahal14}.
From this we can estimate the harmonic coupling parameter of the interaction of the sheet atom
with the substrate $K_1=\omega_0^2M_C=41$~N/m ($M_C$ is the mass of carbon atom).
Therefore, for small displacements, the interaction with the substrate can be described by
the harmonic  potential
\begin{equation}
W(h)=\frac12K_1(h-h_1)^2, \label{f3}
\end{equation}
with stiffness coefficient  $K_1=41$~N/m and equilibrium distance
to the substrate plane $h_1=2.145$~\AA~ \cite{gamo97}.

To describe oscillations of the graphene and h-BN sheet, we present the system Hamiltonian in the
form,
\begin{equation}
H=\sum_{n=1}^{N} \left[\frac12 M_n
(\dot{\bf u}_n,\dot{\bf u}_n)+P_n+\delta_n W(z_n)\right],
\label{f4}
\end{equation}
where $M_n$ is the mass of $n$-th atom of the sheet,
${\bf u}_n=(x_n(t),y_n(t),z_n(t))$ is the radius-vector of $n$-th atom at the time $t$.
The term $P_n$ describes the energy of interaction of the atom with index $n$
with the neighboring atoms, term
$W(z_n)$ -- the energy of interaction of the atom with substrate surface
(the plane of the substrate coincides with the plane $xy$).
Coefficient $\delta_n=1$ if $n$-th atom interacts and $\delta_n=0$
if it does not interact with the substrate
(if it lies above the hole in the substrate).

To describe the dynamics of a graphene sheet, we used the interaction potentials described
in detail in \cite{savin10,savin17}, whereas to describe monolayer hexagonal boron nitride (h-BN) sheet
we used extended Tersoff potential \cite{Los17}.

We consider a hydrogen-terminated graphene (h-BN) sheet, where edge atoms correspond 
to the molecular group CH (BH or NH). 
We consider such a group as a single effective particle at the location of the carbon atom. 
Therefore, in our model of graphene nanoribbons we take the mass
of atoms inside the stripe as $M_n=12m_p$, and for the
edge atoms we consider a larger mass $M_n=13m_p$ (where $m_p=1.6603\cdot 10^{-27}$kg is the proton mass).

If we want to simulate the absence of a substrate for a part of the sheet atoms,
we must take $\delta_n=0$ for these atoms. Figure \ref{fig01} shows a square sheet
of graphene of size $15.6\times15.6$ nm$^2$ consisting of $N=9470$ carbon atoms.
The central circular part of the sheet does not interact with the substrate
(for atoms from this part the coefficient $\delta_n=0$), forming a circular membrane of radius $R=6$~nm.
A similar structure was used to model the vibrations of a circular membrane made of h-BN sheet.
\begin{figure}[tb]
\begin{center}
\includegraphics[angle=0, width=1.0\linewidth]{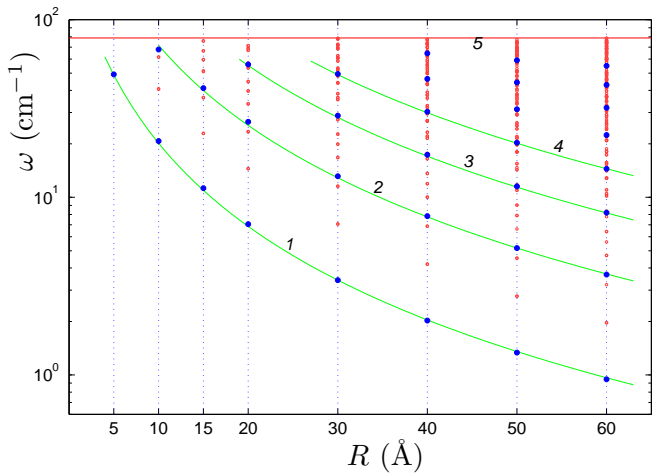}
\end{center}
\caption{\label{fig02}\protect
Dependence of frequencies $\omega$ of intrinsic transversal oscillations of a circular membrane
of a graphene sheet lying on the surface of crystalline h-BN substrate
on the radius of the membrane $R$.
Curves $i=1,2,3,4$ show the dependencies
$\omega_i=c_i/(R+R_i)^2$, where coefficients $c_i=3950$, 15500, 34500, 62500 cm$^{-1}$\AA$^2$,
additional radius $R_i=4.0$, 4.7, 5.0, 5.7~\AA.
Straight line 5 shows the minimum frequency of transversal oscillations
of a graphene sheet on a flat substrate $\omega=\omega_0=78.87$~cm$^{-1}$.
Large markers show natural frequencies of intrinsic nodeless oscillations of the
membrane with circular symmetry.
}
\end{figure}

\section{Transversal normal modes}
Let us consider the transversal vibrations of the atoms of the sheet.
The natural frequencies and normal modes were derived numerically as the solution of the problem
on eigenvalues for matrices of the second derivatives of size $N\times N$.

When only transversal offsets are taken into account, the Hamiltonian of the sheet (\ref{f4})
can be written in the form
\begin{equation}
H=\frac12({\bf M}\dot{\bf Z},\dot{\bf Z})+{\cal P}(\bf Z), \label{f5}
\end{equation}
where ${\bf M}$ is a diagonal matrix of all masses of the sheet, ${\bf Z}=\{z_n-h_0\}_{n=1}^{N}$
is $N$-dimensional vector of transversal displacements from equilibrium positions.
Hamiltonian (\ref{f5}) corresponds to the motion equations,
\begin{equation}
-{\bf M}\ddot{\bf Z}=\frac{\partial}{\partial{\bf Z}}{\cal P}({\bf Z}). \label{f6}
\end{equation}
For small displacements, Eq. (\ref{f6}) reduces to a system of linear equations,
\begin{equation}
-{\bf M}\ddot{\bf Z}={\bf BZ}, \label{f7}
\end{equation}
where the matrix has dimension $N\times N$,
$$
{\bf B}=\left(\left.\frac{\partial^2{\cal P}}{\partial z_{n_1}\partial z_{n_2}}
\right|_{{\bf Z}={\bf 0}}\right)_{n_1=1,~n_2=1}^{N,~N}.
$$

Next, we make the transformation ${\bf Z}={\bf M}^{-1/2}{\bf X}$, and reduce the system (\ref{f7})
to the linear equations of the form $-\ddot{\bf X}={\bf C}{\bf X}$ with the symmetric matrix
${\bf C}={\bf M}^{-1/2}{\bf B}{\bf M}^{-1/2}$. Solutions of this linear system describe
the eigenmodes of the sheet oscillations, which can be presented in the form
${\bf X}(t)=A{\bf e}\exp(i\omega t)$, where $A$ is the oscillation amplitude,
$\omega=\sqrt{\lambda}$ is the frequency, $\lambda$ and ${\bf e}$ are the eigenvalue and normalized
eigenvector of the matrix ${\bf C}$ [$C{\bf e}=\lambda{\bf e}$, $({\bf e},{\bf e})=1$].

The eigenvalues of the matrix ${\bf C}$ can be found numerically.
Numerical matrix diagonalization demonstrates that the presence of the substrate leads to
the presence  of the gap $[0,\omega_0)$ at the bottom of the frequency spectrum of transversal vibrations
(minimum nonzero frequency $\omega_0=\sqrt{K_1/M}$, for graphene $M=M_C$).
All eigen transversal vibrations of the sheet with frequencies $\omega<\omega_0$ correspond
to vibrations localized in the suspended central part of the sheet, i.e. to eigen vibrations
of the circular membrane.

The dependence of the natural oscillations of the circular membrane on its radius is shown in Fig.~\ref{fig02}.
Here, frequency $\omega_0=78.87$~cm$^{-1}$.
A graphene membrane on a crystalline h-BN substrate with a radius of the central hole
in the substrate $R=5$ has only one localized natural oscillation, at $R=10$ -- 6 oscillations,
at $R=15$ -- 14, at $R=20$ -- 23, at $R=30$ -- 55, at $R=40$ -- 94, at $R=50$ -- 144
and at $R=60$~\AA~ -- 209 natural oscillations.
The minimum natural frequency of transversal vibrations of the membrane is approximated
with high accuracy by the dependence
\begin{equation}
\omega_i\sim c_i/(R+R_i)^2, \label{f8}
\end{equation}
with index $i=1$, $c_1=3950$~cm$^{-1}\cdot$\AA$^2$, $R_1=4.0$~\AA.

Asymptotics (\ref{f8}) take place for all substrates, it shows that the interaction of the sheet
with the substrate leads to an additional "increase"\ of the effective radius of the membrane on $R_i$.
The amount of additional magnification depends on the force of interaction with the substrate.
The stronger is the interaction with the substrate, the smaller is the value of $R_1$.
So $R_1=4.7$ for ice substrate having the weakest interaction with a sheet,
$R_1=4.2$ for graphite substrate, $R_1=3.8$ for silicon carbide substrate,
and $R_1=2.1$~\AA \ for substrate with the strongest interaction Ni (111).

Higher frequency natural nodeless oscillations of the membrane with circular symmetry
also have asymptotic (\ref{f8}) with index $i=2$, 3, ... -- see Fig.~\ref{fig02}.
\begin{figure}[tb]
\begin{center}
\includegraphics[angle=0, width=1.0\linewidth]{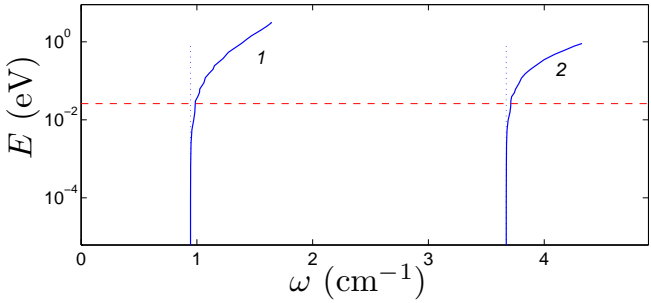}
\end{center}
\caption{\label{fig03}\protect 
Dependence of frequencies $\omega$ of the first and the second (curves 1 and 2)
natural oscillations of the circular membrane ($R=6$~nm) on energy
of vibration $E$.
Energy of thermal vibrations $E=k_BT$ for temperature $T=300$K
is represented by a horizontal dashed line.
}
\end{figure}

\section{Anharmonism of membrane vibrations}
To simulate the natural vibrations of the membrane, we must numerically integrate
a system of equations of motion
\begin{equation}
M_n\ddot{\bf u}_n=-\frac{\partial H}{\partial {\bf u}_n},~~n=1,...,N,  \label{f9}
\end{equation}
with initial conditions
\begin{eqnarray}
x_n(0)=x_n^0,~~y_n(0)=y_n^0,~~,z_n(0)=z_n^0,\nonumber\\
\dot{x}_n(0)=0,~~\dot{y}_n(0)=0,~~\dot{z}_n(0)=Ae_n, \nonumber
\end{eqnarray}
where $\{{\bf u}_n^0=(x_n^0,y_n^0,z_n^0)\}_{n=1}^N$ is the ground state of the graphene sheet,
${\bf e}=\{ e_n\}_{n=1}^N$ is the eigenvector of the matrix ${\bf C}$
(amplitude $A$ determines the energy of vibrations $E=A^2({\bf Me},{\bf e})/2$).
We will use the condition of absorbing edges
(the friction $\Gamma=1/t_r$ with time relaxation $t_r=10$~ps  was introduced 
on edges of the sheet).

Let us consider the dynamics of the natural vibration of the membrane with radius $R=60$~\AA.
Numerical integration of the system of equations of motion (\ref{f9}) has shown that when
energy $E<E_0=0.01$~eV the membrane performs harmonic oscillations with eigen mode frequency
(frequency of the vibration $\omega$ does not depend on the energy $E$).
When $E>E_0$, the frequency of the membrane vibration begins to increase
monotonically when increasing energy -- see Fig.~\ref{fig03}.
Thus, at high energy, the membrane behaves as anharmonic oscillator with rigid anharmonicity.
Let us note that at room temperature $T=300$K, the energy of thermal self-oscilattion of the membrane
$E>E_0$ ($E=k_BT=0.026$~eV).
Therefore, at room temperature, the thermal vibrations of the graphene membrane will be anharmonic.

\section{resonant vibrations}
To analyze the resonant vibrations of a single-layer membrane, we will
simulate the effect of periodic transversal changes in the position of the substrate on its vibrations.
To do this, we numerically integrate the system of Langevin equations of motion
\begin{eqnarray}
M_n\ddot{x}_n&=&-\frac{\partial H}{\partial x_n}+\delta_n[-\Gamma M_n\dot{x}_n+\xi_{n,1}(t)],
\nonumber\\
M_n\ddot{y}_n&=&-\frac{\partial H}{\partial y_n}+\delta_n[-\Gamma M_n\dot{y}_n+\xi_{n,2}(t)],
\label{f10}\\
M_n\ddot{z}_n&=&-\frac{\partial H}{\partial z_n}+\delta_n[-\Gamma M_n\dot{z}_n+\xi_{n,3}(t)+F(t,z_n)],
\nonumber
\end{eqnarray}
where the coefficient $\delta_n=1$ if the atom interacts with the substrate and $\delta_n=0$
if it does not interact (if it is located in the suspended part of the sheet),
$\Gamma=1/t_r$ is the friction coefficient,
and random forces vectors $(\xi_{n,1},\xi_{n,2},\xi_{n,3})$ are normalized as follows:
$$
\langle \xi_{n,i}(t_1)\xi_{m,j}(t_2)\rangle = 2M_n\Gamma k_BT\delta_{nm}\delta_{ij}\delta(t_1-t_2),
$$
$k_B$ -- Boltzmann constant, $T$ -- temperature of the thermostat, $F$ -- force of attraction of the
atom to the substrate.
\begin{figure}[tb]
\begin{center}
\includegraphics[angle=0, width=1.\linewidth]{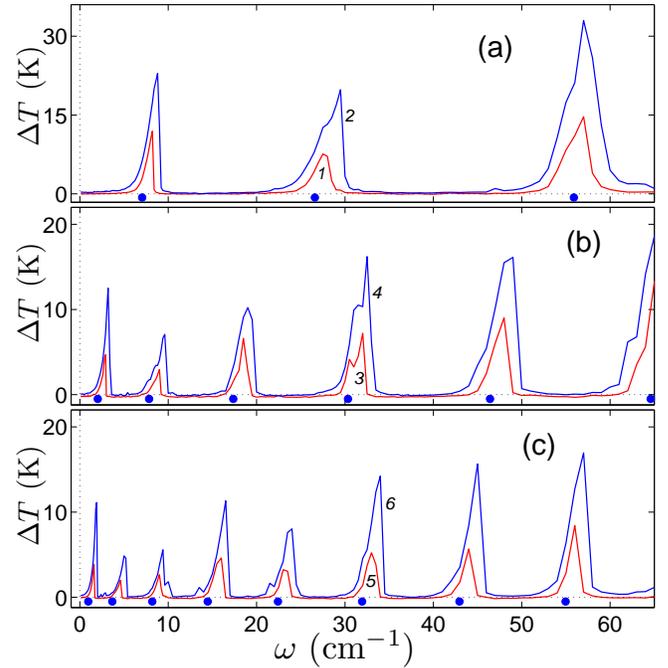}
\end{center}
\caption{\label{fig04}\protect
Dependence of additional thermalization of the graphene membrane $\Delta T$
on the oscillation frequency of the h-BN substrate $\omega$ for (a) membrane radius $R=2$,
(b) $R=4$ and (c) $R=6$~nm.
Red curves 1, 3, 5 show dependencies for amplitude of forced substrate vibrations $A=1$,
blue curves 2, 4, 6 -- dependencies for $A=2$~\AA$\cdot$cm$^{-1}$.
The circular markers show the values of the frequencies of the intrinsic nodeless oscillations
of the membrane with circular symmetry.
}
\end{figure}

If the position of the substrate plane is periodically changed along the $z$ axis,
then in the system of sheet motion equations (\ref{f10}) the force
$$
F(t,z)=-W'(z+\frac{A}{\omega}\cos(\omega t)),
$$
where $A$ and $\omega$ --  the amplitude and the frequency of the forced oscillations of the substrate
(by this definition the amplitude $A$ characterizes the oscillation energy).
\begin{figure}[tb]
\begin{center}
\includegraphics[angle=0, width=1.0\linewidth]{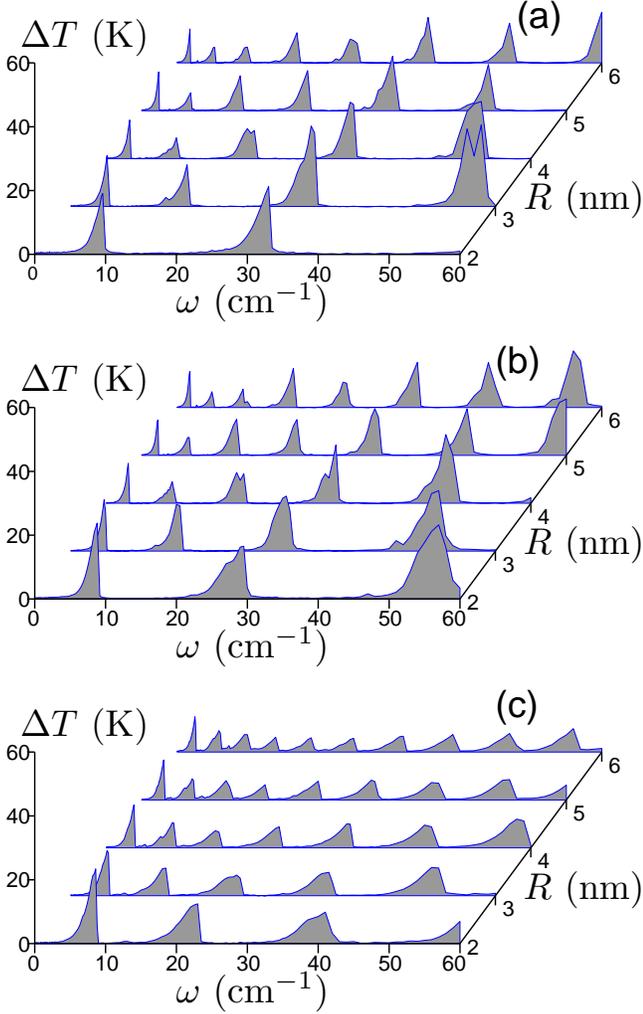}
\end{center}
\caption{\label{fig05}\protect
Dependence of additional thermalization $\Delta T$
on the oscillation frequency of the substrate $\omega$ and on the membrane radius $R$
for graphene membrane on substrates (a) Ni(111), (b) 6H-SiC(0001) and (c) for h-BN membrane
on the substrate 6H-SiC(0001) (amplitude $A=2$~\AA$\cdot$cm$^{-1}$).
}
\end{figure}

Let us analyze at what frequencies of forced oscillations of the substrate 
pumping of the energy to vibrations of the suspended section of the sheet
will be the highest.
For the sheet, the substrate is an external thermostat, so in the system
of the equations of motion (\ref{f10}) only atoms in contact with the substrate interact
with the Langevin thermostat.
The intensity of heat exchange with the thermostat is characterized by a relaxation time $t_r$.
The value $t_r=1$~ps was used in the simulation. In the time $t_0=100t_r$ the sheet being fully thermalized.
The analysis of the further dynamics of the sheet allows us to find the average temperature
of the circular membrane
$$
T_m=\frac{1}{3N_m k_B}\sum_{n=1}^N (1-\delta_n)M_n\langle(\dot{\bf u}_n,\dot{\bf u}_n)\rangle,
~~N_m=\sum_{n=1}^N(1-\delta_n),
$$
where summation occurs only for atoms not in contact with the substrate
($N_m$ is the number of such atoms), and the average value
$$
\langle(\dot{\bf u}_n,\dot{\bf u}_n)\rangle
=\lim_{t\rightarrow\infty}\frac{1}{t}\int_{t_0}^{t_0+t}(\dot{\bf u}_n(\tau),\dot{\bf u}_n(\tau))d\tau~.
$$

When the substrate is stationary (when the oscillation amplitude $A=0$), the temperature
of the membrane is always equal to the temperature of the thermostat ($T_m=T$).
Therefore, additional thermalization of the membrane can be characterized by a temperature difference
$\Delta T=T_m-T$.

Let us take the oscillation amplitude $A=1$, 2~\AA$\cdot$cm$^{-1}$,
the temperature of the thermostat $T=300$K.
The dependence of the additional thermalization of the membrane $\Delta T$ on the frequency
of vertical oscillations of the substrate $\omega$ is shown in Fig.~\ref{fig04}.
As can be seen from the figure, the additional thermalization of the membrane is different from
zero only near certain frequency values, the number of which increases with increasing membrane
radius (see Fig.~\ref{fig05}).
Because by the vertical displacement of the substrate on all the edge atoms of the circular membrane
are the same forces, the vertical vibrations of the substrate in the membrane
can only cause vibrations with circular symmetry.
Therefore, additional thermalization occurs only at frequencies close to the frequencies
of the intrinsic nodeless oscillations of the membrane, which have a circular symmetry
(the amplitude of the displacements of the membrane atom depends only on its distance from
the center of the membrane). Thus, additional thermalization of the membrane occurs primarily
due to the resonant pumping of its own circularly symmetric oscillations.

\begin{figure}[tb]
\begin{center}
\includegraphics[angle=0, width=1.\linewidth]{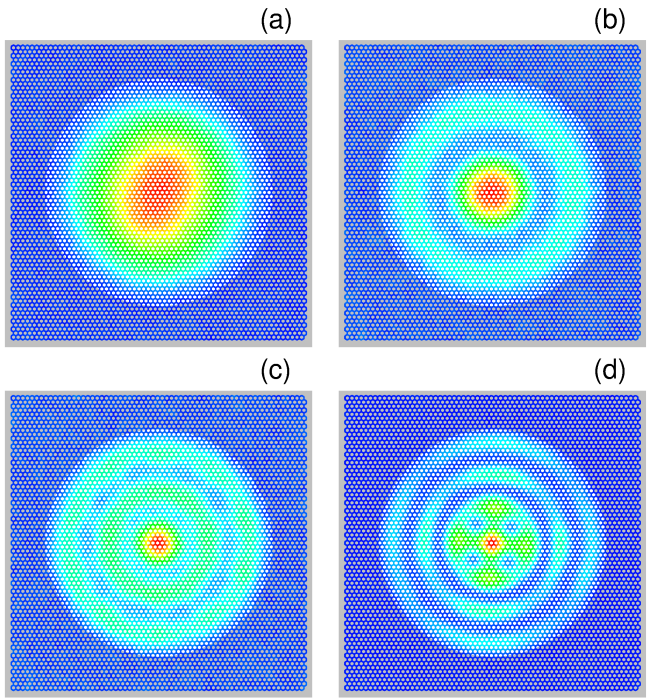}
\end{center}
\caption{\label{fig06}\protect
Temperature distribution in a circular graphene membrane of radius $R=6$~nm
(h-BN substrate, amplitude of forced substrate oscillations $A=2$\AA$\cdot$cm$^{-1}$) at:
(a) first resonance (frequency $\omega=1.9$~cm$^{-1}$, maximum temperature $T_m=331$K);
(b) second resonance ($\omega=5.0$~cm$^{-1}$, $T_m=324$K);
(c) third resonance ($\omega=9.4$~cm$^{-1}$, $T_m=324$K);
(d) the fourth resonance ($\omega=16.5$~cm$^{-1}$, $T_m=366$K).
Blue color corresponds to the background temperature $T=300$K,
red color corresponds to the maximum temperature $T_m$.
}
\end{figure}

A similar resonant pumping of the membrane eigenmodes occurs for the graphene and h-BN sheets
for all considered substrates -- see Fig.~\ref{fig05}.
As can be seen from the figure, the resonant pumping of the main oscillation occurs almost
equally for all membranes.
The differences appear only for higher frequency resonances.
As the membrane radius increases, the resonance frequencies decrease and their number increases.
The analysis of the energy distribution of resonant vibrations of the membrane (see Fig.~\ref{fig06})
shows that the distribution always has a circular symmetry and has several nodal circles
whose number coincides with the number of the resonance frequency.
This shows that resonance pumping occurs primarily due to the excitation of natural oscillations
of the membrane with circular symmetry (i.e. oscillations having only nodal circles).

Let us consider in more detail the first resonance of the membrane.
As can be seen in Fig. \ref{fig04} and \ref{fig05}, when the frequency increases,
the vibrational energy of the membrane initially grows monotonically, at a certain frequency
$\omega_r$ reaches its maximum value, and then sharply decreases to the background value
of the energy of thermal vibrations.
Therefore, it is convenient to determine the frequency of the first resonance as the average value
$$
\bar{\omega}_1=\frac{1}{C}\int_0^{1.1\omega_r}\omega\Delta T(\omega)d\omega,~~
C=\int_0^{1.1\omega_r} \Delta T(\omega)d\omega.
$$
Similarly, we can define the frequencies of next resonances $\bar{\omega}_i$, $i=2$, 3, ....
\begin{figure}[tb]
\begin{center}
\includegraphics[angle=0, width=0.9\linewidth]{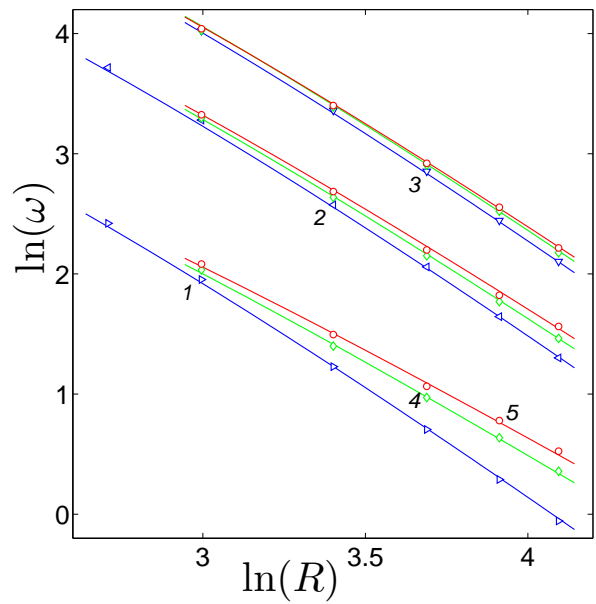}
\end{center}
\caption{\label{fig07}\protect
Dependence of oscillation frequencies $\omega_i$ ($i=1$, 2, 3, markers 1, 2, 3)
and resonance frequencies $\bar{\omega}_i$  on membrane radius $R$
for graphene on h-BN substrate.
Blue curves 1, 2, 3 give approximations $\omega_i=c_i/(R+R_i)^2$,
$c_i=3950$, 15500, 34500~cm$^{-1}\cdot$\AA$^2$,
$R_i=4.0$, 4.7, 5.0~\AA, $i=1$, 2, 3.
Markers 4 give resonance frequencies $\bar{\omega}_i$
for amplitude of forced substrate vibrations $A=1$,
markers 5 -- for $A=2$ \AA$\cdot$cm$^{-1}$.
Green curves give approximations $\bar{\omega}_i=d_i/(R+R_i)^{\alpha_i}$
for $A=1$ ($d_i=1650$, 11900, 32000, $\alpha_i=1.7$, 1.9, 1.96),
red curves -- for $A=2$ ($d_i=1270$, 10500, 2800~cm$^{-1}\cdot$\AA$^{\alpha_i}$,
$\alpha_i=1.6$, 1.85, 1.92).
Dimension of the frequency $[\omega]=$cm$^{-1}$, radius $[R]=$\AA.
}
\end{figure}

The results of numerical simulation of membrane vibrations are shown in Fig.~\ref{fig04}.
The figure shows that each $i$-th eigen membrane vibration with a circular (radial) symmetry
corresponds to resonant membrane vibration with frequency $\bar{\omega}_i>\omega_i$.
The resonance frequency is always higher than the frequency of the corresponding natural
membrane vibration but lower than the frequency of the next natural vibration:
$\omega_i<\bar{\omega}_i<\omega_{i+1}$, $i=1$, 2, 3 ,... .
The larger the amplitude $A$ of forced substrate vibration gets, the stronger
the resonance frequency shifts to the right.
This indicates the nonlinearity of resonances due to rigid anharmonicity of
membrane natural vibration at high energy (the frequency of natural vibration
increases with increasing vibration amplitude).

The analysis of dependency of the resonance frequency $\bar{\omega}_i$ on membrane radius $R$
shows that as the radius increases, the resonance frequency decreases slower than the
frequency of the corresponding natural membrane vibration $\omega_i$:
\begin{equation}
\bar{\omega}_i\sim d_i/(R+R_i)^{\alpha_i},~\alpha_i<2,
\label{f11}
\end{equation}
 -- see Fig.~\ref{fig07}.
The greater the amplitude $A$ of the substrate oscillation, the lower the value of exponent $\alpha_i$.
For the first resonance ($i=1$) the exponent $\alpha_1=1.7$ for $A=1$,
and $\alpha_i=1.6$ for $A=2$ \AA$\cdot$cm$^{-1}$.
The deceleration of the decrease of the  resonance frequencies $\bar{\omega}_i$ with
increasing radius $R$ is caused by the anharmonicity of the membrane vibrations.

\section{Conclusions}
We have simulated natural and resonant oscillations of suspended circular graphene
and hexagonal boron nitride (h-BN) membranes using full-atomic models.
The presence of the substrate (of flat surface of graphite and h-BN crystal, hexagonal ice,
silicon carbide 6H-SiC(0001), nickel surface (111)) leads to the forming of a gap at the bottom
of the frequency spectrum of transversal vibrations of the sheet.
Frequencies of natural oscillations of the membrane $\omega_i$ always lie in this gap,
and they decrease with the increasing radius of the membrane $R$ as $(R+R_i)^{-2}$ with nonezero
effective increase of radius $R_i>0$.
The modeling of the sheet dynamics has shown that small periodic transversal displacements 
of the substrate lead to resonant vibrations of the membranes, at frequencies close 
to the eigenfrequencies of nodeless vibrations of the membranes with circular symmetry.
The energy distribution of the resonant vibrations of the membrane has a circular symmetry
and several nodal circles whose number coincides with the number of the resonant frequency $i$.
The frequencies of the resonances decrease by increasing the radius of the membrane as
$(R+R_i)^{\alpha_i}$ with exponent $\alpha_i<2$.
The lower rate of the resonance frequency decrease is caused by the anharmonicity
of membrane vibrations.

\section*{Acknowledgements}
The author thanks Yuri S. Kivshar for formulating this problem.
This work was supported by the Russian Foundation for
Basic Research, Grant No. 18-29-19135.  
Computational facilities
were provided by the Interdepartmental Supercomputer Center of the Russian Academy of Sciences.

\end{document}